\documentclass[a4paper,aps,pre,groupedaddress,showkeys,showpacs,floats,floatfix]{revtex4-1} 
\usepackage[utf8]{inputenc}
\usepackage[english]{babel}
\usepackage{graphicx}
\usepackage{dcolumn} 
\usepackage{bm}
\usepackage{natbib}
\usepackage{latexsym}
\usepackage{mathrsfs}
\usepackage{amssymb}
\usepackage{amsmath}
\usepackage{amscd}
\usepackage{color}
\usepackage{pifont}
\usepackage{pstricks,pst-node,pst-text,pst-3d}
\usepackage{verbatim}
\usepackage[T1]{fontenc}

\begin{document}  

%\title{Rel}

 \title{  Relativistic virial relation for cosmological structures }
\author{Reza Javadinezhad}
\affiliation{Department of Physics, Sharif University of Technology,
Tehran, Iran }
\email{rezajavadinezhad@gmail.com}

 \author{Javad T. Firouzjaee}
\affiliation{ School of Astronomy, Institute for Research in Fundamental Sciences (IPM), P. O. Box 19395-5531, Tehran, Iran }
 \email{j.taghizadeh.f@ipm.ir}
\author{Reza Mansouri}
\affiliation{Department of Physics and Institute for Advanced Studies, University of Kashan, Kashan, Iran and \\
  School of Astronomy, Institute for Research in Fundamental Sciences (IPM), Tehran, Iran}
 \email{mansouri@ipm.ir- On leave of absence from Department of Physics, Sharif University of Technology.}

%\date{\today}

\begin{abstract}
Starting with the relativistic Boltzmann equation for a system of particles defined by a distribution function, we have derived the virial relation for a spherical structure within an expanding background in the context of general relativity. This generalized form of the virial relation is then applied to the static case of a spherically symmetric structure to see the difference in the simplest case to the Newtonian relation. A relativistic Mass-Temperature relation for this simple case is also derived which can be applied to compact objects in astrophysics. Our general virial relation is then applied to the non-static case of a structure within an expanding universe where an extra term, usually missed in studies of structures in the presence of the dark energy, appears.  
 
\end{abstract}

\maketitle

\section{ introduction}

The study of the virial relation in the context of general relativity has a long history. Beginning with the early work of Lindquist \cite{Lindquist} on the covariant collision-less Boltzmann equation, Jackson \cite{jakson70} published a first formulation of general relativistic virial theorem using the kinetic theory approach for a case of static spherically symmetric structure. His approach has been extended to the virial theorem in the modified gravity theories\cite{virialfr}. There has been another approach to relativistic virial theorem using some of the Einstein equations or its foliation to define energy and derive an equation similar to the classical virial relation. Authors in \cite{virial-rel} derive a relation including terms to be interpreted in the Newtonian limit as potential and kinetic energy using the ADM foliation for a static asymptotically flat space-time, based on the older paper \cite{virial-rel1}. Recently Meyer et al. \cite{lcdm} have written a relativistic version of the virial theorem for the $\Lambda CDM$ cosmologies, using a heuristic approach in defining a potential and kinetic energy term. 
There are different aspects of the context of general relativity, or the formulation of the virial theorem in general relativity. Usually, in speaking of a relativistic condition, one think of the strong gravity regime. There is however another area in the cosmology and astrophysics where general relativistic effects may come in without being in a strong gravity regime. This happens when one is dealing with non-local or large scale phenomena \cite{Buchert-Marra,Jeong-Yoo,Razbin:2012ve,taghizadeh,taghizadeh-mass}. The expansion of 
the universe, as an example, is a general relativistic effect at large scale in a very weak gravity regime: assuming any non-zero density, however small it may be, Einstein equations lead to a non-static expanding universe, in contrast to the Newtonian gravity. The general relativistic virial theorem in the cosmological context, therefore, may reveal novel features not expected from the semi-relativistic static cases. The main motivation of our current study is to find out if one has to expect novel general relativistic features while interested in large scale problems within astrophysics and cosmology.

The paper is organized as follows. In the section II we present a short overview of classical virial relation. The general relativistic version of the virial theorem is discussed in the section III.  We will then apply our relativistic virial relation to some cosmological models in the section IV. The results are then discuss in section V. Appendices A, B and C are aimed
to clarify some of the calculations. Throughout the paper Latin indices are used for orthonormal basis and the Greek ones for the coordinate basis.\\

%%%%%%%%%%%%%%%%%%%%%%%%%% 
\section{classical virial relation}
%%%%%%%%%%%%%%%%%%%%%%%%%%%

Consider a classical system consisting of $N$ particles with the density $\rho$ being bounded by a potential U($\vec{r}$) \cite{book-galaxy formation}. The system is said to be stable if the characteristic time of its evolution is long relative to the characteristic time of the system. The system may then be assumed to be in equilibrium with a distribution function $f$ depending on the phase space coordinates. In case of symmetries, one usually choose coordinates adapted to these symmetries taking care of the distribution function which is a scalar density rather than scalar. With ($\vec{x},\vec{p}$) being the coordinates in the phase space we have
\begin{equation}
N=\int f(\vec{x},\vec{p})~ d^3p ~d^3x.
\end{equation}
Within the kinetic theory of gases, the inertia tensor is then defined as 

\begin{equation}
I_{ij}=\int f x^i x^j d^3p ~d^3x.
\end{equation}
The Euler equation for the system with the pressure $P$, the density
\begin{align}\label{ye1}
\rho=\int m f ~d^3p, 
\end{align}
and the velocity vector field $\vec{v}$, is written as
 \begin{equation}\label{ye2}
 \rho \frac{d\vec{v}}{dt}=-\rho \nabla U -\nabla P.
 \end{equation}

Using now the collision-less constraint, i.e. vanishing the total time derivative of the distribution function, and combining (\ref{ye1}) and (\ref{ye2}) we obtain
\begin{equation}\label{yek}
\frac{\partial f}{\partial t}+ \vec v\frac{\partial f}{\partial \vec x} +\dot{\vec{v}}\frac{\partial f}{\partial \vec v}=0.
\end{equation}

The time derivatives of the inertia tensor, playing a key role in the classical virial theorem, is then given by  
\begin{equation}
\frac{dI_{ij}}{dt}= \int m \frac{\partial f}{\partial t}~x^ix^jd^3p ~d^3x.
\end{equation} 
Using now the constraint equation (\ref{yek}), and integrating by part we obtain 
\begin{equation}
\frac{dI_{ij}}{dt}= \int m f~(v_ix_j+v_jx_i)d^3p ~d^3x,
\end{equation}
where $m$ is the particle mass and $\vec{p}$ is defined as 
\begin{equation*}
\vec{p}=m \vec{v}.
\end{equation*}
The second time derivative leads to the virial relation we are looking for: 
\begin{align*}
\frac{d^2I_{ij}}{dt^2}&=m\int \frac{\partial f}{\partial t}~(v_ix_j+v_jx_i)d^3p ~d^3x \\
									&=m\int ( -\vec v.\frac{\partial f}{\partial \vec x} + (\vec {\nabla} U +\frac{\vec{\nabla }P}{\rho}).\frac{\partial f}{\partial \vec 												p})~(v_ix_j+v_jx_i)d^3p ~d^3x\\
									&=m\int f~(v_kv_i\delta^k_j+v_kv_j\delta^k_i)-m\int f~(\frac{\partial U}{\partial x^k}
									+\frac{1}{\rho}.\frac{\partial P}{\partial x^k})				 (x_i\delta^k_j+x_j\delta^k_i)	d^3p ~d^3x\\		
									&=2m\int f~v_iv_j d^3p ~d^3x -2m\int f~\frac{\partial U}{\partial x^j}	x_i d^3p ~d^3x+2(P_{avg}-P_{b})V	\delta_i^j,																	
\end{align*}
where the average pressure is defined by $ P_{avg}V=\int P~d^3x$. We have added the surface term $2P_b V$, with $P_b$ being the pressure on the boundary of the system, and $V$ the volume of the system. Although it is common to assume that the pressure on the boundary vanishes, we will keep this term for comparison to the relativistic cases. Defining now
\begin{align}
K_{ij}&=\frac{1}{2}\int f~v_iv_j d^3p ~d^3x,\\
W_{ij}&=-\int f~\frac{\partial U}{\partial x^j}x_i d^3p ~d^3x,
\end{align}
we arrive at the following tensorial virial theorem including the surface term:
\begin{align}
&\frac{1}{2}\frac{d^2I_{ij}}{dt^2}=2K_{ij}+W_{ij} +(P_{avg}-P_b) V \delta_i^j.
\end{align}
 The trace of the above equation gives then the classical scalar virial theorem:
\begin{align}
\frac{1}{2}\frac{d^2I}{dt^2}&=2K+W+3(P_{avg}-P_b) V.
\end{align}

Assuming the equilibrium, i.e. $\frac{d^2I}{dt^2}=0$, we arrive at  
\begin{equation}
2K+W+3P_{avg} V=3 P_b V.
\end{equation}
This general classical result is usually the one also used for complex gravitating systems at the solar system and galactic scales. We, however, are looking for cases where the non-linear effects due to general relativity may play a role where one is forced to used the general relativistic effects and is not any more allowed to use this classical virial theorem anymore.\\

\section{relativistic virial relation for non-stationary spherical symmetric space-times}

One of the simplest and widely used cases in astrophysics and cosmology are spherically symmetric structures such as cluster of galaxies. There, it is safe to assume the model of a collision-less gas. Given the expansion of the universe, we therefore assume the space-time metric to be spherically symmetric but in general non-stationary.

%%%%%%%%%%%%%%%%%%%%%%%%%%%%%%%%%%%%%%%%%%%%%%%%%%%%
\subsection{Collisionless Boltzmann Equation}
Let's start with the generalization of the Boltzmann equation in the form (see \cite{Lindquist, boltzman})
%\begin{equation}
%p^\alpha\frac{Df}{dx^\alpha}=C[f] \\ \frac{D}{dx^\alpha}=\frac{\partial}{\partial %x^\alpha}-\Gamma_{\alpha \gamma}^i p^\gamma \frac{\partial}{\partial p^i}. 
%\end{equation}
%or 
\begin{equation}
\frac{df}{d\lambda}=C[f].
\end{equation}
%%????????????????????????????????????????????????????????????????????????????????????????

 The source term C[f] on the RHS takes into account all possible interactions. In the collisionless cases we are interested in this collision term vanishes.

From now on we will use the tetrad basis. This simplifies the calculation specially in the presence of some symmetries where local coordinate systems will be adapted to it. There is a subtlety in calculating the total time derivative where both space-time and momentum coordinates have to be taken into account due to the mass shell relation reducing the number of total variable to 7 (see (\ref{i})). Thus, we may write

\begin{equation} \label{f}
p^aD_aF=0,
\end{equation}
where
\begin{equation}\label{i}
	  D_a=\partial_a-\Gamma_{ac}^i p^c \frac{\partial}{\partial p^i}~~~~~ (i=1,2,3),
\end{equation}
with $\Gamma_{ab}^c$ being the spin connection coefficients in tetrad basis defined by
     \begin{equation}
     \Gamma_{ac}^b\equiv e_a^\alpha e^b_\gamma e^\gamma_{c;\alpha}\label{c}.
     \end{equation}
We now assume a non-static inhomogeneous spherically symmetric space-time given by the Lema\^{i}tre metric in the form 

\begin{align}
&ds^2=-e^{2\alpha(t,r)}~dt^2+e^{2\beta(t,r)}~dr^2+R(t,r)^2d\Omega^2 \label{d}.
%&ds^2=-dt^2+e^{2\beta(t,r)}~dt^2+R(t,r)^2d\Omega^2
\end{align}
Note that we have still the freedom of choosing a new ~$t$~and~$r$~ coordinate such that 
$R$ is independent of $t$ (see appendix C). Using (\ref{e}), (\ref{c}), (\ref{d}), and calculating the coefficients $\Gamma_{ac}^b$, the equation (\ref{f}) leads to 

\begin{align}
&u^i\partial_i f -\frac{\partial f}{\partial u^r}(\alpha' u^t.u^t+\dot{\beta} u^r.u^t-\frac{R'}{R}(u^\theta . u^\theta +u^\phi .u^\phi))-\frac{\partial f}{\partial u^\theta}(\frac{\dot{R}}{R} u^\theta .u^t +\frac{R'}{R} u^\theta .u^r-\frac{\cot(\theta)}{R}u^\phi .u^\phi)\nonumber\\
&-\frac{\partial f}{\partial u^\phi}(\frac{\dot{R}}{R} u^\phi .u^t +\frac{R'}{R} u^\phi .u^r+\frac{\cot(\theta)}{R}u^\theta .u^\phi)=0.
\end{align}
As already noted, although the space time is assumed to be non-static, we still may talk about the equilibrium and the virial relation given that the characteristic time scale of our system is much less than the cosmic time scale. This may be revealed by showing that although the space-time is expanding and having no time-like Killing vector, structures up to about 10 Mpc have a time-like Killing vector within an accuracy of about $10^{-4}$, as shown in the appendix B (see also \cite{lcdm}). This is equivalent to saying the characteristic time scale of the evolution of our system is much larger than the time scale of the expansion of the universe. We are therefore allowed to talk about quasi-equilibrium and the related virial relation.\\
 
 In the following, the derivative with respect to $r$ coordinate is denoted by $'$ while  $\dot{}$ is used to denote the derivative with respect to $t$.
 In our tetrad basis the $r$ dependence due to the spherical symmetry is translated in the dependence on $u^\theta . u^\theta +u^\phi .u^\phi$. We may then write the dependence of $f$ in the form  $f(u^r,(u^\theta)^2+(u^\phi)^2,r,t)$ leading to the condition
 \begin{equation}
 u^\theta \frac{\partial f}{\partial u^\phi} - u^\phi \frac{\partial f}{\partial u^\theta}=0.
 \end{equation}
 
We may then expand the equation (\ref{f}) in the form
\begin{align}\label{g}
&u^t\partial_t f+u^r\partial_r f -\frac{\partial f}{\partial u^r}(\alpha' u^t.u^t+\dot{\beta} u^r.u^t-\frac{R'}{R}(u^\theta . u^\theta +u^\phi .u^\phi))-(\frac{\partial f}{\partial u^\theta} u^\theta +\frac{\partial f}{\partial u^\phi} u^\phi )(\frac{\dot{R}}{R} u^t+\frac{R'}{R} u^r )
=0.
\end{align}
Multiplying now the relation above by $u^r d\omega$ with $d\omega =\frac{dp^r dp^\theta dp^\phi}{p^t}$ being the invariant volume element, and integrating over the momentum space, we arrive at the following form of the Boltzmann equation:

\begin{equation}\label{Vir}
\frac{\partial (\rho \overline{u_t.u_r})}{\partial t}+\frac{\partial (\rho \bar{u}_r^2)}{\partial r}+\alpha(t,r)' \rho (\bar{u}_t^2+\bar{u}_r^2)+(2\dot{\beta}(t,r)+\frac{2\dot{R}(t,r)}{R(t,r)})(\rho \overline{u_t.u_r})
+\frac{R'(t,r)}{R(t,r)}(\rho( 2\overline{u}_r^2-\bar{u}_\theta^2  -\bar{u}_\phi^2))=0.
\end{equation}
The bar over the quantities indicate the average over the momentum space.

%%%%%%%%%%%%%%%%%%%%%%%%%%%%%%%%%%%%%%%%%%%%%%%%%%%%

%%%%%%%%%%%%%%%%%%%%%%%%%%%%%%%%%%%%%%%%%%%%%%%%%%%%%%%%%%%
\subsection{Final Form of the Virial Relation}

Using now the Einstein equations and the energy momentum tensor in the form

\begin{align}
&G_{\rho \nu}=8\pi T_{\rho \nu}
~,~~T_{\rho \nu}=\int f u_\rho u_\nu d\Omega~,~~ 
d\Omega=\frac{du_r~du_\theta ~du_\phi }{u_t},\nonumber
\end{align}
with $G_{01}$ and $G_{00}$  components written as 

\begin{align}
&8\pi \rho(\overline{u_t u_r} )=-2\frac{e^{-(\alpha+\beta)}}{R} [\dot{R'}-\dot{R}\alpha'-\dot{\beta}R']\label{t01},\\
&8\pi \rho(\overline{u_t^2} )=e^{-2\alpha }(\frac{\dot{R}^2}{R^2}+2\frac{\dot{R}}{R} \dot{\beta})+e^{-2\beta}(2\frac{R'\beta'}{R}-2\frac{R''}{R}-\frac{R'^2}{R^2})+\frac{1}{R^2},
\end{align}
the Boltzmann equation (\ref{Vir}) may be written as
\begin{align}\label{25}
&2K + \Omega  \nonumber
-\int R^2 e^{-(\alpha+\beta)}[\ddot{R'}-\ddot{R}\alpha' +\frac{\dot{R}}{R}(\dot{R'}-\dot{R}\alpha' -R'\dot{\beta} +R \alpha' (\dot{\alpha} -\dot{\beta}))+R'(\dot{\alpha}\dot{\beta}-\ddot{\beta}-\dot{\beta}^2)-2\dot{R'}\dot{\alpha}]\nonumber\\&= (4\pi R^3 \rho \overline{u_r^2})\Big\arrowvert_0^{r_0},
\end{align}
where we have used the same symbol $K$ for the following term similar to the Newtonian and static general relativistic kinetic term:
\begin{equation}
K=\frac{1}{2}\int 4\pi R^2R'\rho (\overline{u_r^2}+\overline{u_\theta^2}+\overline{u_\phi^2} ).
\end{equation}
The symbol $\Omega$ stands for 
\begin{equation}\label{26}
\Omega =- \int 4\pi R^3 \alpha' \rho((\overline{u_t^2}+\overline{u_r^2}),
\end{equation}
which in the static case reduces to the familiar potential energy. The third term in this equation is novel and due to the non-static nature of our space-time.

 Note that our systematic approach using a relativistic statistical picture and the Einstein equations leads directly to terms which could be identified as the potential and kinetic energy for a localized structure, in contrast to the ad hoc definitions for the kinetic- or potential energy and the energy conservation based on classical concepts in \cite{virial-rel}, \cite{virial-rel1}, and \cite{lcdm}. In addition, in contrast to the previous papers, our approach is fully embedded in an FRW expanding universe without any approximation, i.e. we have assumed a dynamical exact solution of Einstein equations leading to FRW at infinity which is the most generalized model of a collapsing spherically symmetric cosmological structure within an otherwise FRW universe, without any extra assumptions about the staticity of the solution outside the structure, as in \cite{lcdm}. Therefore, we have implemented the complete non-linear effects of the solution outside the structure on the virial relation within the structure. This assure us of any existing global effect due to the non-linearity of general relativity despite the very weak gravity at large, or due to any local strong gravity effects. Although the assumption of asymptotic flatness is not assumed explicitly in \cite{lcdm}, the mere fact of assuming the Schwarzschild-de Sitter static solution excludes any dynamical effect due to the full FRW expanding solution on the structure and the freezing out of extra degrees of freedom and the non-linearity of the Einstein equations (see \cite {CBH-Rahim-sasaki}).

%%%%%%%%%%%%%%%%%%%%%%%%%%%%%%%%%%%%%%%%%%%%%%%%%%%%%%%%%

\section{application for some cosmological models}

\subsection{Schwarzschild-de Sitter model and its Newtonian Limit}

Let us apply our virial relation (\ref{25}) to a static spherically symmetric space-time, neglecting the back reaction of the system of particles to the metric. Using the Schwarzschild-de Sitter metric form in
\begin{align}
&ds^2=-(1-\frac{2m}{r}+\frac{\Lambda}{3}r^2)dt^2+\frac{1}{1-\frac{2m}{r}+\frac{\Lambda}{3}r^2} ~dr^2+r^2d\Omega^2,
\end{align}
we obtain
\begin{align}
2K + \Omega + \frac{1}{3}\Lambda I =3P_{b}V
\end{align}
as the relativistic virial relation. The symbols $\Omega$ and $I$ are defined as

\begin{align}
\Omega = -\int_{0}^{R_c}\frac{\frac{m}{r}+\frac{\Lambda r^2}{3}}{1-2\frac{m}{r}+\frac{\Lambda r^2}{3}}4\pi\rho r^2dr~,~~
I = \int_{0}^{R_c}4\pi \rho r^4 dr,
\end{align}
similar to the Newtonian counterparts. Note the role of the boundary pressure $P_b$ in the boundary term \cite{surface}:
 \begin{align}
 3P_{b}V = 4\pi \rho_{(R_c)} R_c^3 \overline{u^2_{r(R_c)}}.
 \end{align}
The integrand of $\Omega$ in its general form can be expanded in powers of the Newtonian potential, $\Omega_N=\frac{m}{r}+\frac{1}{6}\Lambda r^2$, used in the familiar structure formation studies at large scales \cite{cosmological-virial}.

\subsection{Virial Theorem inside a star}

One of the simplest application of our approach to the virial theorem is the case of a static structure. This will show us how the general relativistic
modifications of the theorem may looks like. Assuming now a static and spherically symmetric structure, say a star, the TOV equation relating the pressure (p) and the density ($\rho$) is given by 
\begin{equation}
\frac{d p}{d r}=\frac{(\rho+p)(2m(r)+pr^3)}{2r(2m(r)-r)}~~~,~m(r)=\int_{0}^{r}4\pi \rho r^2 dr.\end{equation}
 Our virial relation (\ref{25}) will be simplified by vanishing of the time derivatives of the metric functions. Specifically we need to calculate $\alpha'$ given by the equation of motion
 	\begin{equation}
 		\frac{d p}{d r}=-\alpha' (p+\rho).
 	\end{equation} 
 In addition, assuming 
random local motion of the particles inside the star,  it turns out that for the average of squared component of velocity measured by local observer in 
tetrad frame we have $\overline{u_{\theta}^2}=\overline{u_{\phi}^2}=\overline{u_{r}^2} $. Therefore, the relativistic virial relation for such a configuration 
is given by
\begin{equation}
 2K + \Omega +\int_{0}^{R}\frac{16\pi^2r^4 (p \rho)  dr}{1-\frac{2m(r)}{r}}=  3P_{b} V, 
\end{equation}
with  $\Omega=-\int_{0}^{R}\frac{m(r) dm(r)}{r-
	2m(r)}$. 
An equation of state  for matter is needed to   This is simplified by the assumption of an equation of state $f(\rho, P) = 0$ allowing to eliminate the pressure from the virial equation. Assuming the system is virialized, one can relate the kinetic energy to the temperature \cite{surface} and get the mass-temperature relation. For a monoatomic uniform virialized gas the kinetic energy may be written as
\begin{equation}
K=\frac{3 M_{gas}k_B T}{2 \mu m_p},
\end{equation}
where $k_B$ is the Boltzmann constant, $M_{gas}$ is the total gas mass, $\mu=\rho/n m_p$ is the relative mean molecular mass of the gas, and $m_p$ is the proton mass. Hence the relativistic mass-temperature relation is given by
\begin{align}
T=\frac{\mu m_p} {3 M_{gas}k_B }\left( 3P_{ext} V-\int_{0}^{R}\frac{16\pi^2r^4 (p\rho) dr}{1-\frac{2m(r)}{r}}+ \int_{0}^{R}\frac{m(r) dm(r)}{1-\frac{2m(r)}{r}} \right). 
\end{align}
In astrophysics the mass-temperature (luminosity) relationship is an important issue to verify different models of the star formation \cite{mt-star} and the physics of clusters and galaxies \cite{mt-galaxy}. This relativistic equation may help us to have a more viable verifying relation for   these astrophysical models.
\subsection{Relativistic Virial Relation for Perturbed FRW} 
We turn now to a more realistic example of the structures within an expanding cosmological background. To model such a structure, let us begin with a perturbed FRW model 
representing the dynamics of a structure within the expansing universe in the linear regime\cite{Ma:1995ey}. We use the so-called Newtonian gauge and write the perturbed FRW metric keeping first order terms in the metric functions and velocities ($\overline{u_r^2},\overline{u_\phi^2},\overline{u_\theta^2} \ll \overline{u_t^2} \approx 1$):
\begin{align}
&ds^2=-(1+2\Phi)~dt^2+a(t)^2(1-2\Phi)~(dr^2+r^2d\Omega^2).\\
\end{align}
The virial relation (\ref{25}) is then easily written as
  \begin{equation}\label{q}
  2K + \Omega_F + \Omega_p  -\int a(t)r^2 [\ddot{a}(t)r\Phi' +\dot{a}(t)(\dot{\Phi}+2\frac{\dot{a}(t)}{a(t)} \Phi' r+3 \dot{\Phi'}r)+a(t) \ddot{\Phi'} r]dr=  (4\pi R^3 \rho \overline{u_r^2})\Big\arrowvert_0^{r_0},
 \end{equation}
 where
 \begin{equation}
  \Omega_F = -\int M(R)\frac{dM(R)}{R}
 \end{equation}
 stands for the corresponding FRW potential term. There is an extra potential term due to the perturbed FRW metric:
 \begin{equation}
  \Omega_p = -\int \frac{\int_0^r (\dot{\Phi}(r',t) +H\Phi(r',t))~.(3H a^2r'^2)~dr'}{r} ~dM(R),
  \end{equation}
 where the mass term is defined by $$ M(R)=\int\limits_{0}^{R}4\pi \rho R^2 R' dr,$$ and the physical radius given by $R=a(t) r$. The density of the structure, $\rho$, is related up to the first order of the perturbation potential energy, similar to the Poisson equation, in the following way: 
\begin{align}\label{poisson}
& 4\pi \rho=(a(t)^{-2} \nabla^2 \Phi -3H(\dot{\Phi} +H\Phi)).\\
%&-\int 4\pi R^2R'\rho (\overline{u_r^2}+\overline{u_\theta^2}+\overline{u_\phi^2} )+\int 4\pi R^3 \alpha' \rho-(4\pi R^3 %\rho \overline{u_r^2})\Big\arrowvert_0^{r_0} \nonumber\\
%&-\int a(t)r^2 [\ddot{a}(t)r\Phi' +\dot{a}(t)(\dot{\Phi}+2\frac{\dot{a}(t)}{a(t)} \Phi' r+3 \dot{\Phi'}r)+a(t) %\ddot{\Phi'} r]=0
\nonumber
\end{align} 
 In the weak field limit we arrive again at the classical virial theorem. In the case of vanishing Hubble constant the fourth term in (\ref{q}) will also vanish similar to the static cases already studied.\\
Let's  have a look at the case where the structure is almost static and the time derivatives of the perturbation potential is almost vanishing. Then the virial relation simplifies to
\begin{equation}
 2K + \Omega_F  -\int \frac{\int_0^r (3H^2 a^2r'^2)\Phi(r',t)~dr'}{r} ~dM(R)  -2\int a(t)^2r^3H^2\Phi' (1-\frac{q}{2}) dr=  (4\pi R^3 \rho \overline{u_r^2})\Big\arrowvert_0^{r_0}t.
\end{equation}  
Here is $q=-\frac{a\ddot{a}}{\dot{a}^2}$ the deceleration parameter. The appearance of the third and forth term on the left hand side are due to the cosmological expansion reflecting a novel effect in our virial relation. The third term, corresponding to the last term in the Poisson equation (\ref{poisson}), is a correction to the familiar potential term. The last term is zero for $q=2$. It, however, does not vanish in the accelerated universe with dark energy. Authors using the static form of the de Sitter metric to calculate the cosmological potential correction to dark energy have missed this term \cite{cosmological-virial}.
We expect different implications of these extra novel terms appearing in our general relativistic virial theorem in different astrophysical applications such as  the mass-temperature profile of the galaxy clusters, the Press–Schecter relation for halo mass functions, Sachs-Wolfe and integrated Sachs-Wolfe effect. Simulations like Millennium  or Illustris of large scale structures \cite{illustis} should be an arena to test our virial relation which is beyond the scope of this paper.

%%%%%%%%%%%%%%%%%%%%%%%%%%%%%%%%%%%%%%%%%%%%%%%%%%%%%
\section{discussion}
Using the relativistic distribution function,  we have derived a new form of the general relativistic virial relation for a spherical structure within an expanding universe based on an exact solution of Einstein equations without any ad hoc assumptions. The results are written in the familiar form known from Newtonian gravity. We then have applied it to the case of static spherically symmetric and asymptotically Minkowskian structure, and also within a static star calculating the mass-temperature relation for a relativistic gas model. To be as realistic as possible, we have also applied our virial relation to a structure within a cosmological expanding background (for the relevance and more exact formulation of such structures at large scales see \cite{taghizadeh, taghizadeh-mass, Ma:1995ey}). A new term due to the impact of the expansion of the universe on the structure arise which has been neglected so far in the literature.\\

%%%%%%%%%%%%%%%%%%%%%%%%%%%%%%%%%%%%%%%%%%%%%%%%%%%%%%%%5
\appendix

\section{Einstein Equations}
The non-vanishing Einstein tensor components, assuming the metric (\ref{d}), are $[00,01,11,10,22,33]$. To use them in the virial relation we write them in the tetrad basis. The vierbein for the metric is
\begin{equation} \label{e}
\mathbf{e^a_\rho} = \left(
\begin{array}{cccc}
e^{\alpha(t,r)} & 0&0&0 \\
0& e^{\beta(t,r)} &0&0  \\
0&0&R(t,r)&0\\
0&0&0&R(t,r)\sin(\theta)
\end{array} \right)
,~\mathbf{e^\rho_a} = \left(
\begin{array}{cccc}
e^{-\alpha(t,r)} & 0&0&0 \\
0& e^{-\beta(t,r)} &0&0  \\
0&0&R(t,r)^{-1}&0\\
0&0&0&(R(t,r)\sin(\theta))^{-1}
\end{array} \right).
\end{equation}

To avoid complexity, we use the notations 
\begin{align*}
~R'=\frac{\partial R(t,r)}{\partial r}
~,~\dot{R}=\frac{\partial R(t,r)}{\partial t}.
\end{align*}
The components of the Einstein tensor are
\begin{align}
&G_{01}=G_{10}=-2\frac{e^{-(\alpha+\beta)}}{R} [\dot{R'}-\dot{R}\alpha'-\dot{\beta}R'],\\
&G_{00}=-\frac{e^{-2(\alpha +\beta)}}{R^2}[-2\beta' R R' e^{2\alpha}+2 R R'' e^{2\alpha} -\dot{R}^2e^{2\beta}+R'^2e^{2\alpha} - e^{2(\alpha +\beta)}-2R\dot{R} \dot{\beta} e^{2\beta}],\\
&G_{11}=-\frac{e^{-2(\alpha +\beta)}}{R^2}[2R\ddot{R} e^{2\beta}+ \dot{R}^2e^{2\beta}-R'^2 e^{2\alpha}+e^{2(\beta +\alpha)}-2\dot{\alpha}\dot{R}Re^{2\beta}-2\alpha' R' R e^{2\alpha}],\\
&G_{22}=G_{33}=\frac{e^{-2(\alpha +\beta)}}{R^2}[2RR''e^{2\alpha}+2RR'\alpha'e^{2\alpha}-2RR'\beta'e^{2\alpha}-2R\dot{R} \dot{\beta} e^{2\beta} -2R\ddot{R} e^{2\beta}+2R\dot{R} \dot{\alpha}e^{2\beta}\\ \nonumber
&~~~~~~~~~~~~~~~~~~~~~~~~~~~~~~-2R^2\dot{\beta}^2e^{2\beta}+2R^2\alpha''e^{2\alpha} +2R^2\alpha'^2 e^{2\alpha}-2R^2\ddot{\beta}e^{2\beta}
+2R^2\dot{\alpha} \dot{\beta} e^{2\beta} -2R^2 \alpha' \beta' e^{2\alpha}].
\end{align}

\section{The existence of an approximate time-like Killing vector in the perturbed FRW metric }
 We will argue that it is reasonable to talk about a quasi-static structure within a perturbed FRW universe and to write a virial theorem for it. To this end we show that in a small region around the given structure there is an approximate time-like Killing vector indicating a quasi-equilibrium of the structure allowing to have a virial theorem. Take the perturbed FRW metric and the corresponding Killing equation:
\begin{equation}
ds^2=-(1+2\Phi)~dt^2+a(t)^2(1-2\Phi)~(dr^2+r^2d\Omega^2) ~,~~~L_{\vec{K}}g_{\mu\nu}=0.
\end{equation} 
 Assume the following ansatz as a solution of the above Killing equation:
 \begin{equation}
 \vec{K}=A(r,t)\partial_t+B(r,t)\partial_r.
 \end{equation}
 Inserting this ansatz in the Killing equation, for the time radial component we obtain
 \begin{equation}
 A(\frac{\dot{a}}{a}-\dot{\Phi})+B(\frac{1}{r}-\Phi^\prime)=0.
 \end{equation} 
 Assuming a large galaxy cluster of dimension 10 Mpc,  $\Phi \ll 1$ and $\dot{\Phi} \ll H$, we obtain 
  \begin{equation}
 \frac{B}{A}=-\frac{Hr}{c}\approxeq10^{-4}.
  \end{equation} 
 Therefore, we may talk about a quasi-Killing vector for the structure and a quasi-equilibrium state of it.

 \section{Spherically symmetric metric}

 The general form for a spherically symmetric metric is given by 
 \begin{equation}
 ds^2=-A(t,r)dt^2 + B(t,r) dt dr + C(t,r)dr^2 + D(t,r)(d\theta^2+\sin(\theta) d\phi^2).
 \end{equation}
 We define the new radial variable as $\bar{r}^2=D(t,r)$. This coordinate transformation will change the explicit form of the functions in metric, but since they are arbitrary, the new functions will be denoted as before. The metric may now be written as
  \begin{equation}
 ds^2=-A(t,\bar{r})dt^2 + B(t,\bar{r}) dt d\bar{r} + C(t,\bar{r})d\bar{r}^2 +\bar{r}^2 (d\theta^2+\sin(\theta) d\phi^2).
 \end{equation}
 The cross term can be eliminated by redefinition of the time coordinate. The final form of the general spherically symmetric metric is then written as (skipping the ``bar" over the radial coordinate)
 \begin{equation}
 ds^2=-A(t,r)dt^2 + C(t,r)dr^2 +r^2 (d\theta^2+\sin(\theta) d\phi^2).
 \end{equation}
 Therefore, the most general form of a spherically symmetric metric maybe written, without loss of generality, by two unknown functions.


\begin{thebibliography}{99}
	
\bibitem{Lindquist} 
	R. W. Lindquist, Annals of Physics 37, 487 (1966).
	
	
\bibitem{jakson70} 
	J.~C.~Jackson, Mon. Not. R. Astron. Soc. 148, 249 (1970). 
	
\bibitem{virialfr} 
	T.~Harko and K.~S.~Cheng,
	%``The Virial theorem and the dynamics of clusters of galaxies in the brane world models,''
	Phys.\ Rev.\ D {\bf 76}, 044013 (2007);  
	N.~S.~Santos and J.~Santos,
	%``The virial theorem in Eddington-Born-Infeld gravity,''
	arXiv:1506.04569 [gr-qc].  

\bibitem{virial-rel}
	S. Bonazzola, Astrophys. J. 182, 335 (1973).
	
\bibitem{virial-rel1}
	Eric Gourgoulhon and Silvano Bonazzola, Class. Quantum Grav. 11 443 (1994);
	Silvano Bonazzola and Eric Gourgoulhon, Class. Quantum Grav. 11 1775  (1994).
	
	
	
	
\bibitem{lcdm} 
	S.~Meyer, F.~Pace and M.~Bartelmann,
	%``Relativistic virialization in the Spherical Collapse model for Einstein-de Sitter and \Lambda CDM cosmologies,''
	Phys.\ Rev.\ D {\bf 86}, 103002 (2012).
	
	
\bibitem{taghizadeh}
J.~T.~Firouzjaee and Reza~Mansouri,
%``Asymptotically FRW black holes,''
Gen.\ Rel.\ Grav.\  {\bf 42}, 2431 (2010); W.~Valkenburg,
%``Complete solutions to the metric of spherically collapsing dust in an expanding spacetime with a cosmological constant,''
Gen.\ Rel.\ Grav.\  {\bf 44}, 2449 (2012);
J.~T.~Firouzjaee,
%``The Spherical symmetry Black hole collapse in expanding universe,''
Int.\ J.\ Mod.\ Phys.\ D {\bf 21}, 1250039 (2012); C.~Gao, X.~Chen, Y.~G.~Shen and V.~Faraoni,
%``Black Holes in the Universe: Generalized Lemaitre-Tolman-Bondi Solutions,''
Phys.\ Rev.\ D {\bf 84}, 104047 (2011);
%``Cosmological black holes: the spherical perfect fluid collapse with pressure in a FRW background,''
Class.\ Quant.\ Grav.\  {\bf 32}, no. 21, 215001 (2015);
M.~E.~Rodrigues, M.~H.~Daouda, M.~J.~S.~Houndjo, R.~Myrzakulov and M.~Sharif,
%``Inhomogeneous Universe in f(T) Theory,''
Grav.\ Cosmol.\  {\bf 20}, no. 2, 80 (2014).

\bibitem{Buchert-Marra} 
T.~Buchert,
%``Toward physical cosmology: focus on inhomogeneous geometry and its non-perturbative effects,''
Class.\ Quant.\ Grav.\  {\bf 28}, 164007 (2011); V.~Marra and A.~Notari,
%``Observational constraints on inhomogeneous cosmological models without dark energy,''
Class.\ Quant.\ Grav.\  {\bf 28}, 164004 (2011).

\bibitem{Jeong-Yoo} 
D.~Jeong and F.~Schmidt,
%``Large-Scale Structure Observables in General Relativity,''
arXiv:1407.7979 [astro-ph.CO]; J.~Yoo,
%``Relativistic Effect in Galaxy Clustering,''
Class.\ Quant.\ Grav.\  {\bf 31}, 234001 (2014); C.~Bonvin,
%``Isolating relativistic effects in large-scale structure,''
arXiv:1409.2224 [astro-ph.CO].

\bibitem{Razbin:2012ve} 
  M.~Razbin, J.~T.~Firouzjaee and R.~Mansouri,
  %``Relativistic rotation curve for exact asymtotically FRW spherical structures and the definition of mass,''
  Int.\ J.\ Mod.\ Phys.\ D {\bf 23}, no. 9, 450074 (2014).
  
\bibitem{book-galaxy formation}
Mo, Houjun, Frank Van den Bosch, and Simon White. Galaxy formation and evolution. Cambridge University Press, 2010.

\bibitem{taghizadeh-mass}
J.~T.~Firouzjaee, M.~Parsi~Mood and Reza~Mansouri,
%``Do we know the mass of a black hole? Mass of some cosmological black hole models,''
Gen.\ Rel.\ Grav.\  {\bf 44}, 639 (2012).

\bibitem{boltzman} 
G F R Ellis, R Maartens and M A H MacCallum, \emph{Relativistic
	Cosmology} (Cambridge University Press).

%\bibitem{post-newton}
%S. Chandrasekhar,  Astrophys. J. 142, 1488, (1965).



  
\bibitem{surface} 
A. Del Popolo, 336, 81 MNRAS (2002) ; N.~Afshordi and R.~Cen,
%``Mass - temperature relation of galaxy clusters: A Theoretical study,''
Astrophys.\ J.\  {\bf 564}, 669 (2002).

\bibitem{Ma:1995ey} 
C.~P.~Ma and E.~Bertschinger,
%``Cosmological perturbation theory in the synchronous and conformal Newtonian gauges,''
Astrophys.\ J.\  {\bf 455}, 7 (1995).

\bibitem{cosmological-virial}
Lahav, O., Lilje, P. B., Primack, J. R.,  Rees, M. J. 1991, MNRAS, 251,
128; L. Wang and P. J. Steinhardt, Astrophys. J. 508,
483 (1998).

\bibitem{mt-star}
Hertzsprung, Ejnar. 
%"On the relation between mass and absolute brightness of components of double stars."
 Bulletin of the Astronomical Institutes of the Netherlands 2 (1923): 15; Russell, H. N., W. S. Adams, and A. H. Joy. 
 %"A Comparison of Spectroscopic and Dynamical Parallaxes."
  Publications of the Astronomical Society of the Pacific (1923): 189-193; Baldo, M., et al. 
  %"Proton and neutron superfluidity in neutron star matter."
   Nuclear Physics A 536.2 (1992): 349-365.


\bibitem{mt-galaxy}
Vikhlinin, A., Kravtsov, A., Forman, W., Jones, C., Markevitch,
M., Murray, S. S., \& Van Speybroeck, L. 2006, ApJ, 640, 691.

\bibitem{illustis}
www.illustris-project.org.

\bibitem{CBH-Rahim-sasaki}

R.~Moradi, J.~T.~Firouzjaee and R.~Mansouri,
%``Cosmological black holes: the spherical perfect fluid collapse with pressure in a FRW background,''
Class.\ Quant.\ Grav.\  {\bf 32}, no. 21, 215001 (2015); K.~Yamamoto, V.~Marra, V.~Mukhanov and M.~Sasaki,
%``Perturbed Newtonian description of the Lema\^itre model with non-negligible pressure,''
arXiv:1512.04240 [gr-qc].




\end{thebibliography}
\end{document}